\documentstyle[12pt]{article}
\begin{document}
\begin{titlepage}
\begin{flushright}
{\large \bf  PLB-GG-8192\\ \ \ \\March. 24, 1996}
\end{flushright}
\vspace{0.3in}
\begin{center}
{\LARGE Strong supersymmetric quantum effects on top\\ 
quark production at the Fermilab Tevatron}
\vspace{.5in}
 
{\tt  Chong Sheng Li$^{a,b,e}$, Hong Yi Zhou$^{a,c}$, Yun Lun Zhu$^{a,b,f}$
and Jin Min Yang$^{a,d}$}
 
\vspace{.5in}
 
$^a$CCAST(World Laboratory)\\
P.O.Box 8730, Beijing 100080, P. R. China \\
\bigskip
$^b$ Department of Physics , Peking University,\\
                Beijing 100871, P.R.China\\
$^c$ Department of Physics, Tsinghua University,\\
          Beijing 100084, P.R.China\\ 
$^d$ Physics Department, Henan Normal University,\\
          Xin Xiang, Henan 453002, P.R.China\\
$\mbox{}\hspace{8mm} ^e$ E-mail: csli@pmphpsvr.pku.edu.cn\\
$^f$ E-mail: zhuyl@sun.ihep.ac.cn\\
\vspace{.5in}
 
\end{center}
\begin{footnotesize}
\begin{center}\begin{minipage}{5in}
 
\begin{center} ABSTRACT\end{center}
 
~~~~The supersymmetric QCD corrections to
top quark pair production by $q\bar q$ annihilation in $p\bar p$
 collisions are calculated in the minimal supersymmetric model.
We consider effects of the mixing of the scalar top quarks on
the corrections
to the total $t\bar t$ production cross section at the Fermilab Tevatron.
We found that such correction is less sensitive to squark mass and
gluino mass than in no-mixing case, and in both cases the corrections 
can exceed 10\% even if we consider the recent CDF limit on squark and 
 gluino masses.
\end{minipage}\end{center}
\end{footnotesize}
\vspace{.7in}
 
~~~PACS number: 14.80Dq; 12.38Bx; 14.80.Gt
 
\vspace{0.5in}
 
$^*$ mailing address
 
\vfil
\end{titlepage}
\eject
\rm

\baselineskip=0.35in
 
   The top quark has been discovered by CDF(D0) collaboration at Tevatron[1].
The mass and production cross section are found to be, respectively,
$176\pm 8(stat)\pm 10(syst)\\
(199_{-21}^{+19} (stat) \pm  22 (syst)) {\rm
GeV}$ and $6.8_{-2.4}^{+3.6}(6.4\pm 2.2)pb$ by CDF(D0) collaboration.
At the Tevatron,the dominant production mechanism for a heavy top quark
is the QCD process $q\bar{q}\rightarrow t\bar{t}$[2].
Recently, there has been a lot of interest in the one-loop radiative
corrections of the top quark production cross section at the
Tevatron, which arise from the new physics beyond the SM such as the
two-Higgs-doublet model(2HDM) and the minimal supersymmetric
model(MSSM)[4,5]. In Ref.[5], the supersymmetric QCD corrections to the 
top quark production in $p\bar{p}$ were calculated in the simplest case 
of unmixed squarks and degenerate masses. But we have to know the impact 
from the mixing squarks because of interest to phenomenology, especially, 
sensitive dependence of the supersymmetric QCD on the squark masses and 
gluino mass[5]. In that reference also only two cases of the gluino mass  
$m_{\tilde{g}}=150\,{\rm GeV}$ and $m_{\tilde{g}}=200\,{\rm GeV}$,
respectively, was considered. The purpose of the present letter is to
evaluate the supersymmetric QCD corrections to the top quark production
at the Tevatron in the general case of the mixing squark masses and
compare our results with that in the case of unmixed squark masses given
in Ref.[5]. We also discuss further the dependence of such corrections
on the gluino mass. 
 
In the MSSM the strong supersymmetric interaction Lagrangian relevant
to our calculation is given, in the present of squark mixing, by[6]
\begin{equation}
{\cal 
L}_{\tilde{g}\tilde{q}q}=-i\sqrt{2}g_sT^a(\bar{q}P_R\tilde{q}_L-\bar{q}P_L\tilde{q}_R)\tilde{g}_a+H.C.,
\end{equation}
where $g_s$ is the strong coupling constant, $T^a$ are $SU(3)_c$
generators, $P_{L,R}\equiv \frac{1}{2}(1\pm \gamma_5)$, $\tilde{g}_a$
are the Majorana gluino fields, and $\tilde{q}_{L,R}$ are the current
eigenstate squarks, which are related to the corresponding mass
eigenstates $\tilde{q}_{1,2}$ by
\begin{equation}
\tilde{q}_1 =\tilde{q}_L\cos \theta+\tilde{q}_R\sin \theta,\ \ \ 
\tilde{q}_2=-\tilde{q}_L\sin \theta + \tilde{q}_R\cos \theta  
\end{equation}
The mixing angle $\theta$ as well as the masses $m_{\tilde{q}_{1,2}}$
of the physical squarks are determined by the following mass matrices[7]
\begin{equation}
{\cal M}_{\tilde{t}}^2=\left\{
\begin{array}{c c}
m_{\tilde{t}_L}^2+m_t^2+0.35 D_Z & -m_t(A_t+\mu\cot \beta) \\
-m_t(A_t+\mu\cot \beta) & m_{\tilde{t}_R}^2+m_t^2+0.16 D_Z
\end{array} \right\}, 
\end{equation}
where $D_Z=M_Z^2\cos 2\beta$ with $\tan \beta$ being the ratio 
of Higgs vcauum expectation values, $m_{\tilde{t}_L,\tilde{t}_R}$
are soft breaking masses, $A_t$ is soft breaking parameter describing the
strength of trilinear scalar interactions, and $\mu$ is the supersymmetric
Higgs mass parameter. The relevant Feynman diagrams contributive to one-loop
supersymmetric QCD corrections to $q\bar{q}\rightarrow t\bar{t}$
amplitiude are shown in Fig.1 of Ref.[5]. In our calculation we will follow
the same notation and adopt the same regularization and renormalization
scheme as was used in Ref.[5]. Notice that the off-diagonal elements of
the squark mass matrices are proportional to the quark mass. In the
case of supersymmetric partners of light quarks mixing between the current
eigenstates can therefore be neglected. So we will take into account 
only the mixing between the squarks for the supersymmetric QCD corrections
to the $gt\bar{t}$ coupling and neglect such mixing for the $gq\bar{q}$
couplings because these couplings mainly involved light quarks. The
renormalized amlitude for $q\bar{q}\rightarrow t\bar{t}$ can be written
as
\begin{equation}
M_{\rm ren}=M_0+\delta M_1^{\rm vertex} +\delta M_2^{\rm vertex}+\delta
M^{\rm box}    
\end{equation}
where $M_0$ is the amplitude at tree-level and $\delta M$ represent the
supersymmetric QCD corrections arising from the effective 
$gq\bar{q}(gt\bar{t})$
vertex and box diagrams, which are given by
\begin{eqnarray}
M_0&=&\bar{v}(p_2)(-ig_s T^a 
\gamma^\nu)u(p_1)\frac{-ig_{\mu\nu}}{\hat{s}} 
\bar{u}(p_3)(-ig_s T^a \gamma^\mu)v(p_4) \\  
\delta M_1^{\rm vertex}& = &\bar{v}(p_2)(-ig_s 
T^a\widehat{\Gamma}^{'\nu})u(p_1)
\frac{-ig_{\mu\nu}}{\hat{s}} 
\bar{u}(p_3)(-ig_s T^a\gamma^\mu)v(p_4)\\   
\delta M_2^{\rm vertex}&=&
\bar{v}(p_2)(-ig_s T^a\gamma^\nu)u(p_1)\frac{-ig_{\mu\nu}}{\hat{s}} 
\bar{u}(p_3)(-ig_s T^a\widehat{\Gamma}^\mu)v(p_4)\\  
\delta M^{\rm box} & = & i\frac{7\alpha_sg_s^2}{48
\pi}\{\bar{u}(p_3)P_R u(p_1) \bar{v}(p_2) P_R v(p_4) f_1 \nonumber \\
& & +\bar{u}(p_3)P_R u(p_1) \bar{v}(p_2) P_L v(p_4) f_2
+\bar{u}(p_3)P_L u(p_1) \bar{v}(p_2) P_R v(p_4) f_3 \nonumber \\ 
& & +\bar{u}(p_3)P_L u(p_1) \bar{v}(p_2) P_L v(p_4) f_4 +\bar{u}(p_3)P_R 
u(p_1) \bar{v}(p_2) P_R \not\!{p}_3 v(p_4) f_5 \nonumber \\
& & +\bar{u}(p_3)P_R u(p_1) \bar{v}(p_2) P_L \not\!{p}_3 v(p_4) f_6 
+\bar{u}(p_3)P_L u(p_1) \bar{v}(p_2) P_R \not\!{p}_3 v(p_4) f_7 \nonumber \\
& & +\bar{u}(p_3)P_L u(p_1) \bar{v}(p_2) P_L \not\!{p}_3 v(p_4) f_8 
+\bar{u}(p_3)\not\!{p}_4 P_R u(p_1) \bar{v}(p_2) P_R  v(p_4) f_9 
\nonumber \\
& & +\bar{u}(p_3)\not\!{p}_4 P_R u(p_1) \bar{v}(p_2) P_L  v(p_4) f_{10} +
\bar{u}(p_3)\not\!{p}_4 P_L u(p_1) \bar{v}(p_2) P_R  v(p_4) f_{11} 
\nonumber \\
 & & +\bar{u}(p_3)\not\!{p}_4 P_L u(p_1) \bar{v}(p_2) P_L  v(p_4) f_{12} +
\bar{u}(p_3)\not\!{p}_4 P_R u(p_1) \bar{v}(p_2) P_R \not\!{p}_3 
v(p_4)f_{13} \nonumber \\ 
 & &+\bar{u}(p_3)\not\!{p}_4 P_R u(p_1) \bar{v}(p_2) P_L \not\!{p}_3 
v(p_4) f_{14} +
\bar{u}(p_3)\not\!{p}_4 P_L u(p_1) \bar{v}(p_2) P_R \not\!{p}_3 v(p_4) 
f_{15} \nonumber \\
 & & +\bar{u}(p_3)\not\!{p}_4 P_L u(p_1) \bar{v}(p_2) P_L \not\!{p}_3 
v(p_4) f_{16} +
\bar{u}(p_3)\gamma_\mu P_R u(p_1) \bar{v}(p_2) P_R \gamma^\mu v(p_4) 
f_{17} \nonumber \\
 & & +\bar{u}(p_3)\gamma_\mu P_R u(p_1) \bar{v}(p_2) P_L \gamma^\mu 
v(p_4) f_{18} +
\bar{u}(p_3)\gamma_\mu P_L u(p_1) \bar{v}(p_2) P_R \gamma^\mu v(p_4) 
f_{19} \nonumber \\
 & & +\bar{u}(p_3)\gamma_\mu P_L u(p_1) \bar{v}(p_2) P_L \gamma^\mu 
v(p_4)f_{20} \}
\end{eqnarray} 
with
\begin{eqnarray}
\widehat{\Gamma}^{' \nu} & = & \gamma^\nu F_1^{'}+\gamma^\nu \gamma_5
F_2^{'} + k^\nu F_3^{'}+k^\nu \gamma_5F_4^{'}+ik_\mu\sigma^{\nu\mu}F_5^{'}+
ik_\mu\sigma^{\nu\mu}\gamma_5F_6^{'}\\ 
\widehat{\Gamma}^\mu & = & \gamma^\mu F_1+\gamma^\mu \gamma_5F_2+
k^\mu F_3+k^\mu \gamma_5F_4+ik_\nu\sigma^{\mu\nu}F_5+
ik_\nu\sigma^{\mu\nu}\gamma_5F_6 
\end{eqnarray}
Here,$k=p_1+p_2\ $,$p_1,p_2$ denote the momenta of incoming partons,and
$p_3,p_4$ are used for outgoing top-quark and its antiparticle.
$F_i^{'},F_i$ and $f_i$ are form factors which are presented in
Appendix. 
 
The renormalized differential cross section of the subprocess is given by
\begin{equation}
\frac{d\hat{\sigma}}{d\cos\theta}=\frac{1}{32\pi}\frac{\beta_t}{\hat{s}}
\vert M_{\rm ren}\vert^2, 
\end{equation}
where
\begin{equation}
\vert M_{\rm ren}\vert^2=\overline{\sum}\vert M_0\vert^2
+2{\rm Re}\overline{\sum}M_0^+ (\delta M_1^V+\delta M_2^V+\delta M^b), 
\end{equation}
with
\begin{equation}
\overline{\sum}\vert M_0\vert^2 = \frac{4g_s^4}{9\hat{s}^2}[2\hat{s}m_t^2+
(\hat{t}-m_t^2)^2+(\hat{u}-m_t^2)^2],
\end{equation}
\begin{equation}
\overline{\sum}M_0^+\delta M_1^V = \frac{4g_s^4}{9\hat{s}^2}[2\hat{s}m_t^2+
(\hat{t}-m_t^2)^2+(\hat{u}-m_t^2)^2](F_1^{'}-1),
\end{equation}
\begin{equation}
\overline{\sum}M_0^+\delta M_2^V = \frac{4g_s^4}{9\hat{s}^2}\{[2\hat{s}m_t^2+
(\hat{t}-m_t^2)^2+(\hat{u}-m_t^2)^2](F_1-1) 
 + 2m_t^2\hat{s}^2F_5\},
\end{equation}
\begin{eqnarray}
\overline{\sum}M_0^+\delta M^b& =& 
\frac{7g_s^4}{216\hat{s}}\frac{\alpha_s}{\pi}
\{[\hat{s}m_t^2+(\hat{t}-m_t^2)^2](f_2+f_3)\nonumber\\
 & & -m_t[\hat{s}^2 + \hat{s}\hat{t} - 2m_t^2\hat{s} 
-(\hat{t}-m_t^2)(\hat{s}+\hat{t}-m_t^2)]
(f_6+f_7-f_{10}-f_{11})\nonumber\\ 
& & -[m_t^4\hat{s}+(\hat{s}-m_t^2)(\hat{s}+\hat{t}-m_t^2)^2+
2m_t^2(\hat{t}-m_t^2)\nonumber\\ 
& &  
(\hat{s}+\hat{t}-m_t^2)](f_{14}+f_{15})
-2[m^2_t\hat{s}+(\hat{s}+\hat{t}-m^2_t)^2](f_{18}+f_{19})\}, 
\end{eqnarray}
In the above equations, $\theta$ is the scattering angle between the quark
and the top quark, $\hat{s}$, $\hat{t}$, and $\hat{u}$ are the kinematic 
invariants for
the $2\rightarrow 2$ subprocess with $\hat{s}+\hat{t}+\hat{u}=2m_t^2$,
and $\beta_t=\sqrt{1-4m_t^2/\widehat {s}}$ the velocity of the final quarks.
The total cross section for the production of top quark pair can be written
in the form
$$
\sigma (s)=\sum_{i,j}\int dx_1dx_2\widehat{\sigma }_{ij}(x_1x_2s, m_t^2, 
\mu^2)
[F_i^A(x_1,\mu)F_j^B(x_2,\mu)+(A \leftrightarrow B)], \eqno(17)
$$
with
$$
\begin{array}{ll}
s=(P_1+P_2)^2, & \widehat{s}=x_1x_2s,\\
p_1=x_1P_1, & p_2=x_2P_2,
\end{array} \eqno(18)
$$
where A and B denote the incident hadrons, $P_1$ and $P_2$ are their
four-momenta, $i,j $ are the initial partons, $x_1$ and $x_2$ are their
longitudinal momentum fractions, and the functions $F^A_i$, $F^B_j$ are
the parton distributions of the initial-state hadrons A and B.
In our numerical calculations, we have used
the MRS Set $A^\prime$ parton distribution functions $[8]$, and do not
consider SUSY corrections to the parton distribution functions since the
principle of decoupling demands that the corrections are negligible.
Introducing a convenient variable $\tau=x_1x_2$ and changing to $x_1$
and $\tau$ as independent variables, the total cross section expression
becomes
$$
\sigma (s)=\sum_{i,j}\int\limits_{\tau_0}^{1}\frac{d\tau}{\tau}(\frac{1}{s}
\frac{dL_{ij}}{d\tau})(\widehat{s}\widehat{\sigma_{ij}}), \eqno(19)
$$
where $\tau_0=4m_t/s$ and the quantity $dL_{ij}/d\tau$ is the parton 
luminosity,
which is defined as
$$
\frac{dL_{ij}}{d\tau}=\int\limits_{\tau}^{1}\frac{dx_1}{x_1}[F_i^A(x_1,\mu),
F_j^B(\frac{\tau}{x_1},\mu) + (A \leftrightarrow B)]. \eqno(20)
$$
 
  Now we present some numerical examples. 
In our numerical calculation, we input $m_t=176$ GeV
and 1-loop $\alpha_s(Q^2=\hat{s})$ and use the phase space cuts:
$|\eta|<2.5,\;\;p_T>20 GeV$.
As to the supersymmetric parameters involved in our calculations,
in general, once $\tan\beta$ and $m_{\tilde t_L}$ are fixed, we are
free to choose two independent parameters in the stop mass matrix
: $(m_{\tilde t_R}, A_t+\mu\cot\beta)$, which also can be tranfered 
to $(m_{\tilde t_R},m_{\tilde t_1})$. 
As explained above, we neglect the mixing of squark masses in
the corrections to the $gq\bar{q}$ couplings and also neglect the mass 
splittings between squarks of different flavor for simplicity.
 
  In Figs.1-3 we give some numerical results in a simple case,
in which we set $\tan\beta=1$ and $m_{\tilde t_L}=m_{\tilde t_R}
 =m_{\tilde t_1}=m_{\tilde q}$ and thus the mixing angle is $\pi/4$ and
$A_t+\mu=m_t$.  To compare the results in no-mixing case with those in
the mixing case, we show the results in both cases in Figs.1-3.
From these figures we can see that the corrections in mixing case
are smaller than in the no-mixing case.
The plots versus squark and gluino masses in the mixing case get smoother 
than in no-mixing case.
Fig.1 show the dependence of the corrections on gluino mass for fixed 
squark 
mass of 150 GeV. The corrections can be either positive or negative, 
depending on the gluino mass.  The corrections are negative for gluino 
mass 
below 150 GeV and above that they become positive. When gluino mass is 
changed
from 100 GeV to 200 GeV, the corrections vary from -2\%(-10\%) to 
16\%(23\%) 
in mixing (no-mixing) case. The corrections get their positive maximum 
size at 
gluino mass of 200 GeV. When gluino mass is larger than 200 GeV,
the corrections in both cases drop monotonously with the increase of 
gluino 
mass and tend to zero at 600 GeV, showing the decoupling effects. 
The recent CDF lower limit [9] on gluino mass is 160 GeV  for arbitrary 
squark mass and 220 GeV when gluino mass is equal to squark mass. 
So the corrections are always positive if we consider the CDF limit on 
gluino mass. 
Fig.2 and Fig.3 show the dependence of the corrections on squark mass
for fixed gluino masses of 150 GeV and 200 GeV, respectively.
The corrections in the mixing case differ significantly
from  those in no-mixing case for low squark mass.
For gluino mass of 200 GeV
the CDF lower limit [9] for squark mass is about 220 GeV. At this lower
limit the corrections are 14\% and 18\% for the mixing and no-mixing cases,
respectively. But with the increase of squark mass the difference of the 
corrections in both cases is getting negligibly small.
  
  From Fig1-3 we have found that the corrections vary rapidly with gluino
mass, especially for 150 Gev $\leq m_{\tilde g}\leq$ 200 GeV, though 
mixied  
cases are smaller than unmixied ones. This is due to the fact that we 
have 
set $m_t=176$GeV in the numerical calculation, and the threshold for open 
top quark production is crossed in that region. If we change the top 
quark 
mass, we can find that such region is also shifted correspondingly, 
which  
provides a check on our calculation, especially on the treatment of the  
threshold.  
  
  We also perform the numerical calculations for $\tan\beta=10$. We found
that the corrections are not sensitive to $\tan\beta$ value. For example,
with gluino mass of 200 GeV and squark mass of 150 Gev we get 
$\Delta\sigma=0.482\,pb$ for $\tan\beta=1$ and $\Delta\sigma=0.490\,pb$ 
for 
$\tan\beta=10$.       
  
  In conclusion, we have shown that the supersymmetric QCD corrections to
top quark pair production by $q\bar q$ annihilation in $p\bar p$
collisions can exceed 10\% in both the mixing and no-mixing cases of 
squark masses even if we consider the recent CDF limit on squark and 
gluino masses.
 
\bigskip
We thank J. Ellis for suggesting this calculations.
This work was supported in part by the National Natural Science Foundation
of China.
 
\vspace{0.3in}
\begin{center}{\large Appendix } \end{center}
The form factor $F_1$ is given by
$$
F_1=1+\frac{\alpha_s}{3\pi}\left [ A_{ii}\left ( F_1^{(t\tilde g \tilde t_i)}
        +2m_t^2 G_1^{(t\tilde g \tilde t_i)}\right )\right.
$$
$$
\left. +2m_tm_{\tilde g}B_{ii} G_0^{(t\tilde g \tilde t_i)}
        +\frac{1}{8}(F_1^{(1)}+9F_1^{(2)})\right ] ,\eqno(21)
$$
where
$$
A_{ii}=a_ia_i+b_ib_i,~~B_{ii}=a_ia_i-b_ib_i,\eqno(22)
$$
$$
a_1=-b_2=\frac{1}{\sqrt 2}(\cos\theta-\sin\theta),
a_2=b_1=-\frac{1}{\sqrt 2}(\cos\theta+\sin\theta),\eqno(23)
$$
$$
F_1^{(t\tilde g \tilde t_i)}=\int_0^1 dy ~y\ln \left [
        \frac{-m_t^2y(1-y)+m^2_{\tilde g}(1-y)+m^2_{\tilde t_i}y}{\mu^2}
        \right ],\eqno(24)
$$
$$
G_n^{(t\tilde g \tilde t_i)}=-\int_0^1 dy ~
        \frac{y^{n+1}(1-y)}{-m_t^2y(1-y)+m^2_{\tilde g}(1-y)+m^2_{\tilde 
t_i}y}
        ,\eqno(25)
$$
$$
F_1^{(1)}=2m_t[m_{\tilde g}B_{ii}(c_0+c_{11})-m_tA_{ii}(c_{11}+c_{21})]
        -2A_{ii}\bar{c}_{24}(-p_3,k,m_{\tilde g},m_{\tilde t_i},
m_{\tilde t_i}),\eqno(26)
$$
$$
F_1^{(2)}=-1+A_{ii}[2\bar{c}_{24}+\hat s (c_{22}-c_{23})-m^2_{\tilde g}c_0
-m_t^2(c_0+2c_{11}+c_{21})](-p_3,k,
m_{\tilde t_i},m_{\tilde g},m_{\tilde g})
$$
$$
-2B_{ii}m_tm_{\tilde g}(c_0+c_{11})(-p_3,k,
m_{\tilde t_i},m_{\tilde g},m_{\tilde g}),\eqno(27)
$$
Here $\theta$ and $\theta'$ are the mixing angels of squarks.
$c_0, c_{ij} $ are the 3-point Feynman integrals, definition and expression
for which  can be found in Ref.[10].
The form factor $F'_1$ is obtained by
$$
F'_1=1+\frac{\alpha_s}{3\pi}\left [ K_{jj}F_1^{(q\tilde g \tilde q_j)}
        +\frac{1}{8}(F_1^{(1)'}+9F_1^{(2)'})\right ],\eqno(28)
$$
where
$$
K_{jj}=A_{jj}\vert_{\theta\rightarrow \theta'},
L_{jj}=B_{jj}\vert_{\theta\rightarrow \theta'},\eqno(29)
$$
$$
F_1^{(1)'}=-2K_{jj}\bar{c}_{24}(p_1,-k,m_{\tilde g},m_{\tilde q_j},
m_{\tilde q_j} ),\eqno(30)
$$
$$
F_1^{(2)'}=-1+K_{jj}[2\bar {c}_{24}+\hat s (c_{22}-c_{23})-m_{\tilde g}^2c_0]
(p_1,-k,m_{\tilde q_j},m_{\tilde g},m_{\tilde g}) \eqno(31)
$$
$F_5$ is given by
$$
F_5=\frac{\alpha_s}{24\pi}\left [ F_5^{(1)}+9F_5^{(2)} \right ],\eqno(32)
$$
where
$$
F_5^{(1)}=[m_tA_{ii}(c_{11}+c_{21})-m_{\tilde g}B_{ii}(c_0+c_{11})]
        (-p_3,k,m_{\tilde g},m_{\tilde t_i},m_{\tilde t_i} ),\eqno(33)
$$
$$
F_5^{(2)}=[m_tA_{ii}(c_{11}+c_{21})+m_{\tilde g}B_{ii}c_{11}]
        (-p_3,k,m_{\tilde t_i},m_{\tilde g},m_{\tilde g}),\eqno(34)
$$
$f_i$ are given by
$$
f_2=m^2_{\tilde g}\sigma^{\prime 2}_{ij}D_0
        -m_tm_{\tilde 
g}\sigma_{ij}\sigma^{\prime}_{ij}(D_{12}+D_{13}-D_0-D_{11})
        +m_t^2\sigma^2_{ij}(D_{26}-D_{12}-D_{24}),\eqno(35)
$$
$$
f_6=-m_{\tilde g}\sigma_{ij}\sigma^{\prime}_{ij}(D_0+D_{11})
        +m_t\sigma^2_{ij}(D_{12}+D_{24}),\eqno(36)
$$
$$
f_{10}=m_{\tilde g}\sigma_{ij}\sigma^{\prime}_{ij}(D_{13}-D_{12})
-m_t\sigma^2_{ij}(D_{12}+D_{23}+D_{24}-D_{13}-D_{26}-D_{25}),\eqno(37)
$$
$$
f_{14}=\sigma^2_{ij}(D_{12}+D_{24}-D_{13}-D_{25}), \eqno(38)
$$
$$
f_{18}=\sigma^2_{ij}D_{27}, \eqno(39)
$$
$$
\left (f_3,f_7,f_{11},f_{15},f_{19}\right )
=\left (f_2,f_6,f_{10},f_{14},f_{18}\right )
        \left \vert_{\sigma_{ij}\rightarrow \lambda_{ij}}\right. \eqno(40)
$$
where
$$
\sigma_{ij}=(a_i+b_i)(c_j+d_j), \sigma'_{ij}=(a_i-b_i)(c_j+d_j),
$$
$$
\lambda_{ij}=(a_i-b_i)(c_j-d_j), \lambda'_{ij}=(a_i+b_i)(c_j-d_j),
$$
$$
(c_j,d_j)=(a_j,b_j)\vert_{\theta\rightarrow \theta')} \hfill \eqno(41)
$$
Here $D_0, D_{ij}(-p_1,-p_2,p_4,m_{\tilde g},m_{\tilde q_j},
m_{\tilde g},m_{\tilde t_i})$ are the 4-point Feynman integrals [10].
 
\vfill
\eject
 
\vspace{0.2in}
{\LARGE References}
\vspace{0.3in}
\begin{itemize}
\begin{description}
 
\item[{\rm[1]}] CDF Collaboration, Fermilab-Pub-94/097-E (1994).
\item[{\rm[2]}] F.Berends, J.Tausk and W.Giele, Phys.Rev.D47, 2746(1993).
\item[{\rm[3]}] P.Nason, S.Dawson and R.K.Ellis, Nucl.Phys.B303,607(1988);\\
                 G.Altarelli et al.,Nucl.Phys.B308,724(1988);\\
                 W.Beenakker et al., Phys.Rev.D40, 54(1989);\\
 E.Laenen, J.Smith and W.L.van Neerven, Phys.Lett.B321,254(1994);\\
 W. Beenakker et al., Nucl.Phys.B411,343(1994);\\
 E.L.Berger and H.Contopanagos, Preprint, Argonne ANL-HEP-PR-95-31.
\item[{\rm[4]}] A Stange and S.Willenbrock, Phys.Rev.D48,2054(1993).
\item[{\rm[5]}] J.M.Yang and C.S.Li, Phys.Rev.D52,1541(1995);\\
                C.S.Li, B.Q.Hu, J.M.Yang and C.G.Hu, Phys.Rev.D52, 
5014(1995).
\item[{\rm[6]}] H. E. Haber and C. L. Kane, Phys. Rep. 117, 75 (1985);
                    J. F. Gunion and H. E. Haber, Nucl. Phys. B272, 1 
(1986).        
\item[{\rm[7]}] J.Ellis and S.Rudaz, Phys.Lett.B128,248(1983);\\
                M.Drees and M.M.Nojiri, Nucl.Phys.B369,54(1992);\\ 
                A.Djoudi, M.Drees and H.Konig, Phys.Rev.D48, 3081(1993).
\item[{\rm[8]}] A.D.Martin,W.J.Stirling and R.G.Roberts, Phys. Lett B 145,
                155(1995).
\item[{\rm[9]}] CDF collaboration, Fermilab-CONF-95/172-E,
                to appear in the proceedings of the 10th workshop
                         on proton-antiproton collisions by Jay Hauser.
\item[{\rm[10]}] G. Passarino and M. Veltman, Nucl. Phys. B160, 151 (1979);
                 A. Axelrod, Nucl. Phys. B209, 349 (1982);
                 M. Clements et al.,  Phys. Rev. D27, 570 (1983).
 
\end{description}
\end{itemize}
\vfil
\eject
 
\begin{center}
{\large Figure Captions}
\end{center}
Fig. 1 ~~  Relative correction to hadronic cross section at Tevatron
versus gluino mass.\\
Fig. 2 ~~Same as Fig.1, but versus squark mass for gluino mass of 150 GeV.\\
Fig. 3 ~~Same as Fig.2, but for gluino mass of 200 GeV.\\
 
\end{document}